\documentclass{INTERSPEECH2023}

\interspeechcameraready

\title{The DeepZen Speech Synthesis System for Blizzard Challenge 2023}
\name{Christophe Veaux$^1$, Ranniery Maia$^1$, Spyridoula Papandreou$^1$}
\address{$^1$DeepZen, UK}
\email{christophev@deepzen.io, rannierym@deepzen.io, spyridoulap@deepzen.io}

\begin{document}

\maketitle

\setlength{\abovedisplayskip}{8pt}
\setlength{\belowdisplayskip}{8pt}

\begin{abstract}
This paper describes the DeepZen text to speech (TTS) system for Blizzard Challenge 2023. The goal of this challenge is to synthesise natural and high-quality speech in French, from a large monospeaker dataset (hub task) and from a smaller dataset by speaker adaptation (spoke task). We participated to both tasks with the same model architecture. Our approach has been to use an auto-regressive model, which retains an advantage for generating natural sounding speech but to improve prosodic control in several ways. Similarly to non-attentive Tacotron, the model uses a duration predictor and gaussian upsampling at inference, but with a simpler unsupervised training. We also model the speaking style at both sentence and word levels by extracting global and local style tokens from the reference speech. At inference, the global and local style tokens are predicted from a BERT model run on text. This BERT model is also used to predict specific pronunciation features like schwa elision and optional liaisons. Finally, a modified version of HifiGAN trained on a large public dataset and fine-tuned on the target voices is used to generate speech waveform. Our team is identified as O in the the Blizzard evaluation and MUSHRA test results show that our system performs second ex aequo in both hub task (median score of 0.75) and spoke task (median score of 0.68), over 18 and 14 participants, respectively.

\end{abstract}

\section{Introduction}
The Blizzard Challenge\footnote{https://www.synsig.org/index.php/Blizzard\_Challenge} is a yearly occurring shared task aiming at advancing TTS by comparing and understanding different approaches, with extensive evaluation by human listeners. This year’s challenge consists of
a “hub” task and a “spoke” task. For both tasks, participants are required to build a voice for synthesising French texts. 

\begin{itemize}
\item \textbf{Hub task}: to build a voice from about 50h of data from a French female speaker, training the model using only publicly available data.
\item \textbf{Spoke task}: to build a voice from about 2h of data from another French female speaker, by using speaker adaptation of a pre-trained model. There are no requirements on the use of external data for training the initial model. 
\end{itemize}

We participated to both tasks with a model architecture derived from our current production models. Our design choices were led by two aims: achieving the best possible acoustic and prosodic naturalness, while allowing prosodic control at both global (utterance) and local (word) levels.

The naturalness of speech produced by TTS systems has significantly improved with neural-based techniques. A common approach consists of a two-stage architecture where an acoustic model predicts a mel-spectrogram from the linguistic input (phonemes with punctuation marks) and a neural vocoder synthesises speech waveform from it. In this paper we mainly focus on the acoustic model. Autoregressive acoustic models using an attention-based encoder-decoder, such as Tacotron2~\cite{shen2018natural}, achieve high naturalness but suffer from robustness issues and lack of fine-grained prosodic control. Non-autoregressive acoustic models such as FastSpeech2 \cite{ren2022fastspeech} or Fastpitch \cite{fastpitch} offer greater  robustness and better prosodic control as they use duration and pitch predictors. However, auto-regressive decoders still have an intrinsic advantage for generating consistent prosody at the frame level and thus more natural sounding speech. Therefore we adopt an architecture similar to non-attentive Tacotron (NAT) \cite{shen2021nonattentive} with a duration predictor and gaussian upsampling but modify it to allow simpler unsupervised training. More specifically, we use a simple gaussian attention with strict monotonic property, similar to \cite{tian2020feathertts}, to train the duration predictor and gaussian upsampling. At inference, the attention module is bypassed and the model relies only on the duration predictor and gaussian upsampling.

The two-stage TTS architecture using mel-spectrogram as intermediate representation introduces a training/inference gap since the vocoder is exposed to ground-truth mel-spectrograms at training but predicted mel-spectrograms during inference. Recently, single-stage end-to-end architectures \cite{kim2021conditional, tan2022naturalspeech, casanova2023yourtts} have been introduced to generate speech waveform directly from the linguistic input and avoid the training/inference gap. However, this gap can also be greatly reduced by fine-tuning the vocoder on the synthesised mel-spectrograms and this is the approach followed in this paper.

As predicting prosody from text is a one-to-many problem, there has been considerable work in learning latent style representations for better prosodic modeling and controllability. Initial methods \cite{wang2018style, zhang2019learning} obtained latent representation at the sentence level by encoding the target mel-spectrogram into a fixed-length embedding. As sentence-level representation is not enough to model fine-grained prosodic variability, more recent works use the alignment information to aggregate mel-spectrogram or other prosodic features at the phone or word levels and extract localised prosody embeddings \cite{klimkov2019finegrained, sun2020fullyhierarchical}. It is difficult however to learn disentangled representations in a fully unsupervised manner \cite{locatello2019challenging}. One solution is to estimate the finer-level latent conditioned on the coarser-level as in \cite{sun2020fullyhierarchical}. We follow a different approach which uses distinct features for the sentence and word levels, with the word-level feature being normalised over the sentence level. At the sentence level, a reference encoder extracts a global embedding from the mel-spectrogram as in \cite{wang2018style}. At the word level, we consider the pitch contour mean- and variance-normalised over the sentence and decomposed into a pitch spectrogram by a continuous wavelet transform (CWT) \cite{suni2016cwt}. A secondary reference encoder is used to extract a sequence of local  embeddings from this pitch spectrogram, at the onsets of the words as given by the alignment. In this way, the local reference embeddings encode information such as pitch prominence, whereas the global reference embedding encodes information such channel conditions, vocal quality, global pace, pitch register and excursion. Similarly to \cite{wang2018style}, the global and local reference embeddings are fed to separate multi-head attentions with trainable codebooks of style tokens as keys and values in order to produce global and local style embeddings, respectively. While reference mel-spectrogram and normalised pitch contour are used at training to extract global and local style embeddings, we need to select the appropriate style embeddings from text only at inference. Following an approach inspired by \cite{karlapati2020prosodic}, we use a BERT model to predict the global and local style embeddings from the input text. Additionally, we use pre-trained speaker embedding to model speaker information. 

Finally, our text frontend is able to generate variants of pronunciations. In French, most of these pronunciation variants correspond to schwa elision and optional liaisons. At training, we use a lattice-based forced-alignment that selects the correct pronunciation variant from the recordings. The forced-alignment is also used to classify the pauses into broad categories (short, medium, long), which are transcribed in the linguistic input as "categorical pause" markers. We found out that these categorical pause markers helps to stabilise the attention mechanism used to train the duration predictor and gaussian upsampling\footnote{Because of this, the training of our model may be seen as semi-supervised rather than fully unsupervised.}. At inference, the same BERT model as above is used to predict the pronunciations variants (schwa elision and optional liaisons) and the categorical pauses. 
The key elements of our approach can be summarised as:
\begin{itemize}
\item Unsupervised NAT training with single gaussian attention.
\item Prosodic variability modeled by global and local style tokens, together with pre-trained speaker embeddings.
\item BERT model used to predict the global and local style embeddings, as well as pronunciation variants and categorical pauses.
\end{itemize}

\section{Acoustic Model}

\subsection{Overview}
\begin{figure*}[!t]
  \centering
  \begin{subfigure}{.5\textwidth}
  	\centering
  	\includegraphics[width=\linewidth]{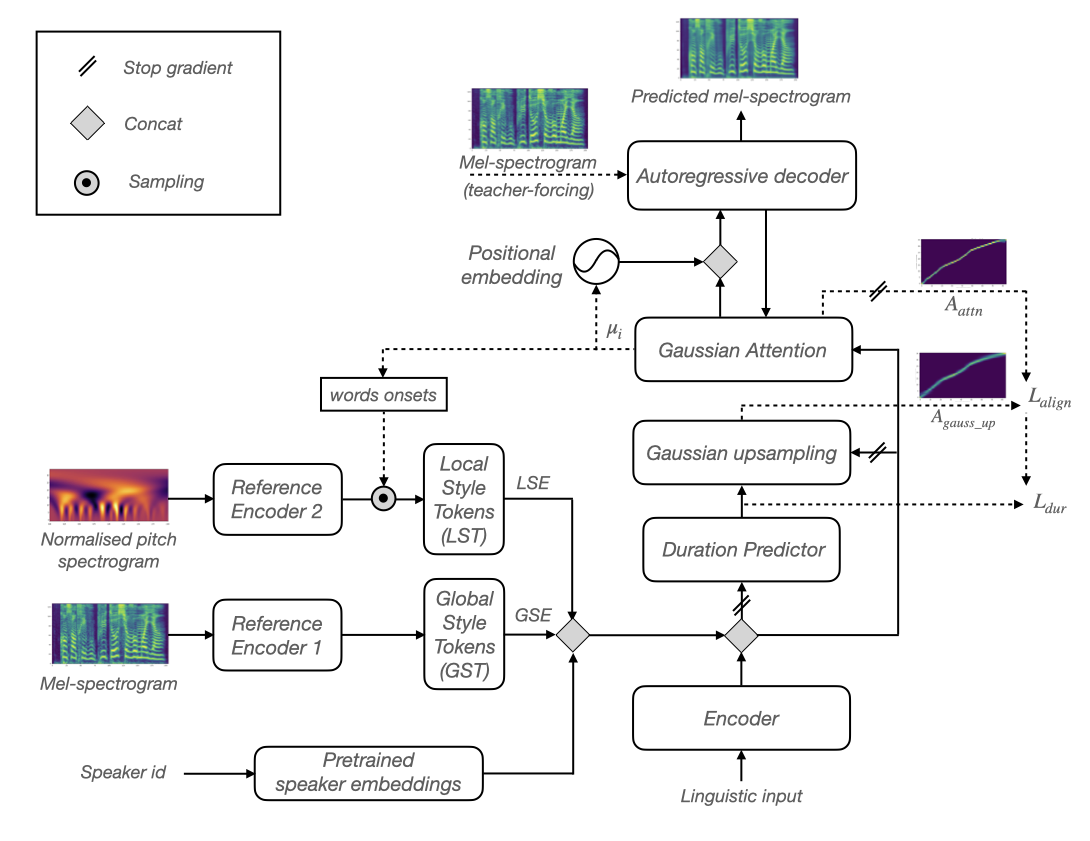}
  	\caption{Training}
  \end{subfigure}%
  \begin{subfigure}{.5\textwidth}
  	\centering
  	\includegraphics[width=\linewidth]{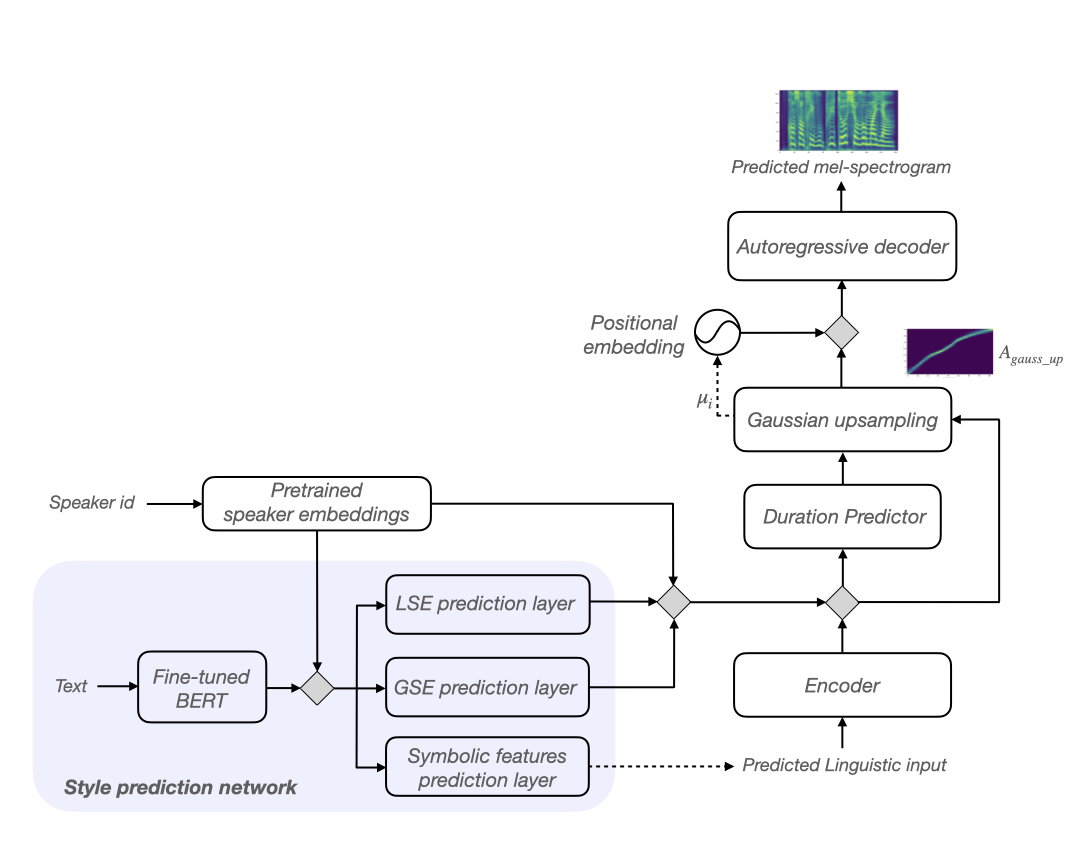}
  	\caption{Inference}
  \end{subfigure}
  \caption{Abstract diagram of the acoustic model at training and inference}
  \label{fig:acoustic_model}
\end{figure*}

The schematic architecture of the acoustic model is shown in Figure~\ref{fig:acoustic_model} for training (a) and inference (b). At training, a gaussian attention learns the alignment $A_{attn}$ between the target mel-spectrogram and the encoder outputs. From this alignment, we estimate the durations of each encoder outputs. These durations and the alignment are fed as targets for the duration predictor and the gaussian upsampling, respectively. The alignment is also used to sample the local reference embeddings at the words onsets, in order to create word-level local style embeddings LSE. These local style embeddings are concatenated with the global style embedding GSE and the speaker embedding at the output of the encoder. Since the decoder is autoregressive, estimating the LSE at the start of the words allows to use them to condition the generation of each word. The architecture of the encoder and decoder follows that of Tacotron \cite{shen2018natural} except that we use layer normalisation instead of batch normalisation in the PostNet and add a projection layer after it. We detail the key components of the model in the following subsections.

\subsection{Attention-driven training}
Most TTS systems with explicit duration model are trained in supervised fashion with "ground-truth" durations extracted from external sources such as forced-alignment or attention-based TTS models. However, an unsupervised approach is more practical and consistent because the durations that we need to model are those of the encoder outputs, and external alignments not necessarily optimally represent those. An unsupervised learning objective was proposed in \cite{badlani2021tts} that uses the forward-sum algorithm to maximise the likelihood over all possible monotonic alignments. This objective makes the assumption of independence between alignment steps which is valid for parallel TTS models but doesn't hold for auto-regressive models. Therefore, in the auto-regressive case, it reduces to an additional loss term encouraging monotonicity but not enforcing it. Instead, we propose here a modified gaussian attention to ensure locality, monotonicity and completeness of the alignment. Similarly to \cite{tian2020feathertts}, we model the alignment at step $i$ with a single gaussian:
\begin{gather}
	\alpha_{i,j} = \exp(-\frac{(j-\mu_i)^2}{2\sigma_i^2}) \\
	\mu_i = \mu_{i-1} + \Delta_i
\end{gather}
where the parameters $(\sigma_i, \Delta_i)$ are non-linear transforms of intermediate parameters $(\hat{\sigma_i}, \hat{\Delta_i})$ calculated from the decoder hidden state $h_i$:
\begin{gather}
	(\hat{\sigma_i}, \hat{\Delta_i}) = V\,\mathrm{Relu}(W h_i + b) \\
	\sigma_i = \mathrm{Softplus}(\hat{\sigma_i}) \\
	\Delta_i = \mathrm{Sigmoid}(\hat{\Delta_i})
\end{gather}
The sigmoid transform constrains the alignment to be monotonic and non-skipping, which is a desirable property in order to derive encoder outputs durations. The mean $\mu_i$ gives the location of the attention at decoding step $i$ and can be mapped to the position of the words. Furthermore, using a purely location-based attention allows to dynamically concatenate the LSE associated with a given word to the encoder outputs once we start to attend this word, since the attention mechanism doesn't use the encoder outputs as precomputed keys.

\subsection{Duration prediction, Upsampling, Positional encoding}
The architecture of the duration predictor and gaussian upsampling follows \cite{shen2021nonattentive} with some simplifications. The duration predictor passes the encoder outputs\footnote{In this section we denote as encoder outputs, the concatenation of the encoder outputs with the GSE, LSE and speaker embedding.} through a single bi-directional LSTM layer followed by a projection layer to predict the log-durations of each encoder output. These are compared to the log-durations derived from the attention-based alignment $A_{attn}$ via the L2 loss term $L_{dur}$. For the gaussian upsampling, the predicted durations are passed through a 1-D convolutional layer to project them to an embedding space. The resulting duration embeddings are concatenated with the encoder outputs and passed through a projection layer and a SoftPlus activation to predict the range parameter $\sigma$ for each encoder output \cite{shen2021nonattentive}. The soft alignment $A_{gauss\_up}$ produced by the gaussian upsampling is compared to the attention-based alignment $A_{attn}$ via the KL-divergence loss term $L_{align}$. The minimisation of the losses $L_{dur}$ and $L_{align}$ only impacts the weights of the duration predictor and gaussian upsampling networks since we stop the gradient flow to the encoder outputs and to the gaussian attention network. Finally, slightly differing from the NAT implementation \cite{shen2021nonattentive}, we encode the fractional progression of the alignment $\mu_i$ with respect to the current encoder output via a Transformer-style sinusoidal positional embedding concatenated at the input of the decoder. We found this positional embedding to be important for the stability of the inference mode, when the duration predictor and gaussian upsampling are used instead of the attention module, and where $\mu_i$ is derived as the centroid at decoding step $i$ of the soft alignment $A_{gauss\_up}$ as shown in Figure~\ref{fig:acoustic_model} (b).

\subsection{Global and Local Style Layers}
Two parallel style layers extract the global and local style embeddings, GSE and LSE, from the reference speech. As in \cite{wang2018style}, the embedding spaces are learned as a convex combination of trainable style tokens. The architecture of the global style layer follows \cite{wang2018style} with a reference encoder consisting of a 2-D convolutional stack followed by a GRU network whose last state is used as query for the global style tokens layer to generate the global style embedding GSE. For the local style layer, the GRU network is replaced by a bi-directional GRU and we consider the outputs at each time-step. Using the alignment location $\mu_i$, we sample these outputs at the time-steps corresponding to the words onsets to form a sequence of queries for the local style tokens layer and generate the local style embeddings LSE. The input of the local style layer is the continuous wavelet transform (CWT) of the pitch contour normalised over the sentence. We use the Wavelet Prosody toolkit\footnote{https://github.com/asuni/wavelet\_prosody\_toolkit} to extract a smoothed pitch contour, interpolated over unvoiced gaps, normalise it over the sentence and apply the CWT. Finally, the speaker characteristics are encoded via a separate embedding generated by the \textit{advanced\_gru\_network} of \cite{ruggiero2021voice} which we trained on a combination of LibriTTS~\cite{zen2019libritts}, VCTK~\cite{VCTK2017} and DAPS~\cite{daps2015} datasets. 

\section{Style Prediction Network}
At inference, a separate style prediction network conditioned on both text and speaker information predicts the GSE at the sentence level, and the LSE as well as some symbolic linguistic features at the word level. However, rather than predicting directly the GSE and LSE, we found beneficial to predict the combination weights over the global and local style tokens, from which we derive the predicted GSE and LSE. This ensures that the predicted embeddings are within the embedding spaces learned by the global and local style layers. A pre-trained BERT model \cite{devlin2019bert} extracts contextualised word-piece embeddings from the text and these are concatenated with the speaker embedding. Following an architecture similar to \cite{karlapati2020prosodic}, the resulting embeddings are passed through a bidirectional LSTM and the first and last hidden states are concatenated to get the GSE combination weights. For the LSE combination weights and other word-level features, the first word-piece embedding of each words, concatenated with the speaker embedding, are simply passed through an affine layer, similarly to \cite{talman2019predicting}.

To train the GSE and LSE prediction layers, we use the global and local style layers of Figure~\ref{fig:acoustic_model} (a) to extract the GSE and LSE over all training sentences. As in \cite{stanton2018predicting}, we use the cross-entropy loss between predicted combination weights and target attention weights for the GSE and LSE.

The word-level linguistic features predicted by the style network are the pronunciation variants (schwa elision, optional liaisons) and a 5-way classification of the normalised pause durations\footnote{For this categorisation, the pause durations extracted by the forced-alignment are normalised by the average speaking rate of the speaker.} denoted as "categorical pauses". When training the acoustic model, these categorical pauses are estimated by forced alignment and inserted in the phonemes and punctuations sequence at the input of the encoder to help stabilise the attention. The forced alignment is also used to select the correct pronunciation variants over the training sentences. For training the symbolic features prediction layer, we use the cross-entropy loss between the annotations given by the forced alignment and the predicted symbolic outputs. However, since the categorical pauses are naturally ordered, we add a regularisation term to their loss, based on the squared Earth Mover's Distance \cite{hou2017squared}.

\begin{figure}[h]
  \centering
  \begin{subfigure}{.15\textwidth}
  	\centering
  	\includegraphics[width=\linewidth]{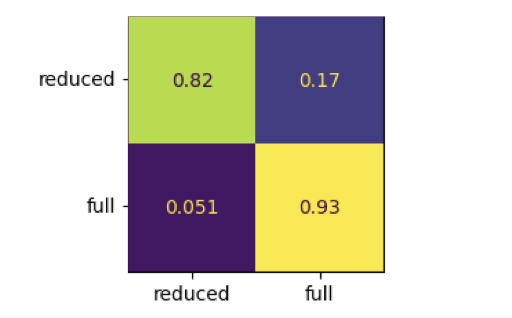}
  	\caption{schwa elision}
  \end{subfigure}%
  \begin{subfigure}{.15\textwidth}
  	\centering
  	\includegraphics[width=\linewidth]{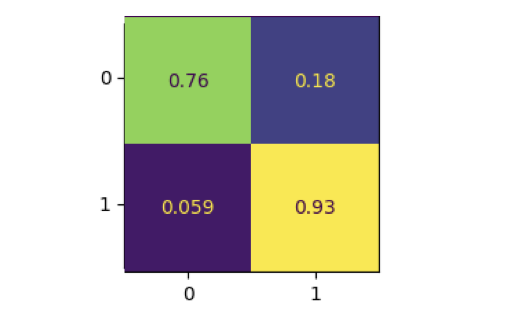}
  	\caption{optional liaison}
  \end{subfigure}
   \begin{subfigure}{.2\textwidth}
  	\centering
  	\includegraphics[width=\linewidth]{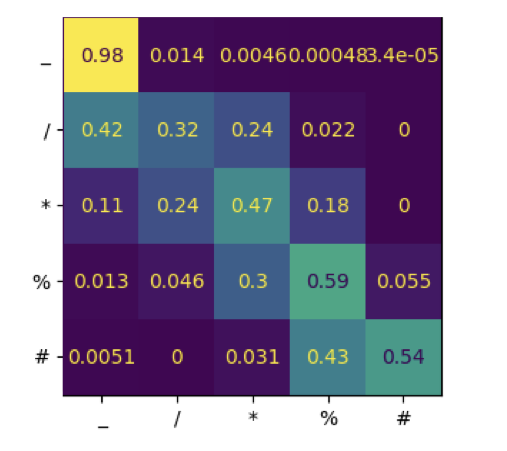}
  	\caption{categorical pauses}
  \end{subfigure}%
  \caption{Confusion matrices of the predicted symbolic features (ground-truth value is on ordinate and predicted on abscissa)}
  \label{fig:prosody_net}
\end{figure}

Finally, the losses from the GSE, LSE and symbolic features prediction layers are combined in an overall loss used to fine-tune the BERT model. The confusion matrices of the predicted symbolic features for the French female speaker of the hub task are shown in Figure~\ref{fig:prosody_net}, with full/reduced and 1/0 denoting the realisation/omission of the schwa and optional liaison, respectively. The normalised pauses durations are classified in an ordinal fashion from "no pause" (-) to "extra-long" (\#).

\section{Data Processing, Vocoder and Training}

\subsection{Text processing}
We leverage our internal text frontend based on Spacy~\cite{honnibal2020spacy} to convert the French texts into sequences of phonemes. We added to the Spacy pipeline our own custom components to perform rule-based text normalisation and basic homograph disambiguation, using part-of-speech (POS) contexts, as well as phonetisation. For the phonetic transcription, we first perform a lexicon lookup using a large lexicon with phonetic variants derived from \cite{lexique3}. Then out-of-vocabulary words are converted to phonemes with an attention-based sequence-to-sequence G2P trained on that same lexicon. Finally, a set of post-lexical rules are applied to enforce mandatory liaisons. When used to process training data, this pipeline allows to generate phonetic variants associated mainly to schwa elision and optional liaisons. These variants are used by the Montreal-Forced Aligner (MFA)~\cite{McAuliffe2017MontrealFA} to decode the correct phonetic transcription. At synthesis, the style prediction network is first run on text to decide the phonetic variants depending on the speaker as well as to predict the categorical pauses which are inserted in the phonetic sequence generated by the text pipeline.

\subsection{Vocoder model}
We use a modified version of HiFiGAN~\cite{kong2020hifigan} that we pre-trained on the same datasets (LibriTTS, VCTK, DAPS) than the speaker encoder. First, compared to HiFiGAN-v1, we add one more stack of convolution layers and residual connection to each ResBlock of the generator in order to capture longer contexts. Second, in the multi-scale discriminator, we replace the average pooling by max-pooling as it helps reducing metallic artefacts in unvoiced segments. As for our mel-spectrogram, we use 128 bins ERB-scale frequency wrapped spectrograms computed on 50 ms frame size with 10 ms hop size.

\subsection{Training Strategy}
The overall loss function for the acoustic model is
\begin{gather}
	L = L_{spec} + \lambda(L_{dur} + L_{align})
\end{gather}
where $L_{spec}$ is the combined L1 loss between ground-truth and predicted mel-spectrograms before and after the PostNet. We use Adam optimiser with batch size 32 and a learning rate of $10^{-3}$ exponentially decaying after 50K steps, $\lambda$ is set to 0 until 50K and then to 1.

The style prediction network is trained separately using the pre-trained uncased Flaubert model \cite{le2020flaubert}. This model is fine-tuned over the combination of the GSE, LSE and symbolic features prediction losses. We use Adam optimiser with a learning rate of $2\mathrm{e}{-5}$ and batch size 8.

Both models are trained on the following datasets:
\begin{itemize}
\item \textbf{Hub task}: 50h from target French female speaker (NEB), "book" and "parl" subsets of the Siwis dataset~\cite{Honnet2017TheSF} (about 10h from another French female speaker).
\item \textbf{Spoke task}: besides the hub task datasets, we use extra internal dataset of about 15h (3 French female speakers). 
\end{itemize}
For the spoke task, once the initial acoustic and style prediction models are trained, we fine-tune them on the unseen target speaker data (AD). The vocoder model is always fine-tuned to the target speakers for both hub and spoke tasks. Finally, the embeddings dimensions for the acoustic model are given in Table~\ref{tab:embedding_dim}, otherwise stated similar to \cite{shen2018natural}.
\begin{table}[th]
  \caption{Embedding dimensions for the acoustic model.}
  \label{tab:embedding_dim}
  \centering
  \begin{tabular}{ c c }
    \toprule
   speaker embeddings & 256~~~             \\
   global / local style embedding & 256~~~			\\
   global / local style tokens & 32~~~			\\
   num global / local style tokens & 10~~~			\\
   positional embedding & 32~~~			\\
   duration embedding & 32~~~			\\
   gaussian attention hidden size & 128~~~			\\
    \bottomrule
  \end{tabular}
\end{table}

We train our acoustic model on a 32GB NVIDIA V100 GPU while HifiGAN and BERT fine-tuning are done on a 16GB V100 GPU. Inference is performed on 2.70GHz Intel Xeon Platinum CPU, where the end-to-end synthesis (including the text frontend) runs 3.8 times faster than real-time in average.

\section{Blizzard Evaluation}
A total of 18 teams participated in the hub task, and 14 in the spoke task. Our system is identified with letter O. The letter A
denotes the natural speech from the original speaker. Two baseline systems were added, a FastSpeech2 + HiFi-GAN benchmark denoted as BF, and a grapheme-based Tacotron2 + HiFi-GAN benchmark denoted as BT. 

The initial evaluation for the hub task includes: naturalness test with MOS (mean opinion score), similarity test with SMOS (similarity mean opinion score), SUS intelligibility test with WER (word error rate), and homographs pronunciation, which is not included here as we focus on the acoustic model. For the spoke task, the initial evaluation comprises naturalness MOS and similarity SMOS. All the systems were evaluated in these initial tests, however the baseline models BF and BT scored very low, which somewhat compresses the scale for the best models. Therefore, a subset of the best 3-4 models was selected to perform a MUSHRA quality test for both tasks. Our system was selected for these MUSHRA tests and ranked 2nd ex aequo for both hub and spoke tasks. 

\subsection{Hub task}
\begin{figure}[h!]
  \centering
  \includegraphics[width=0.9\linewidth]{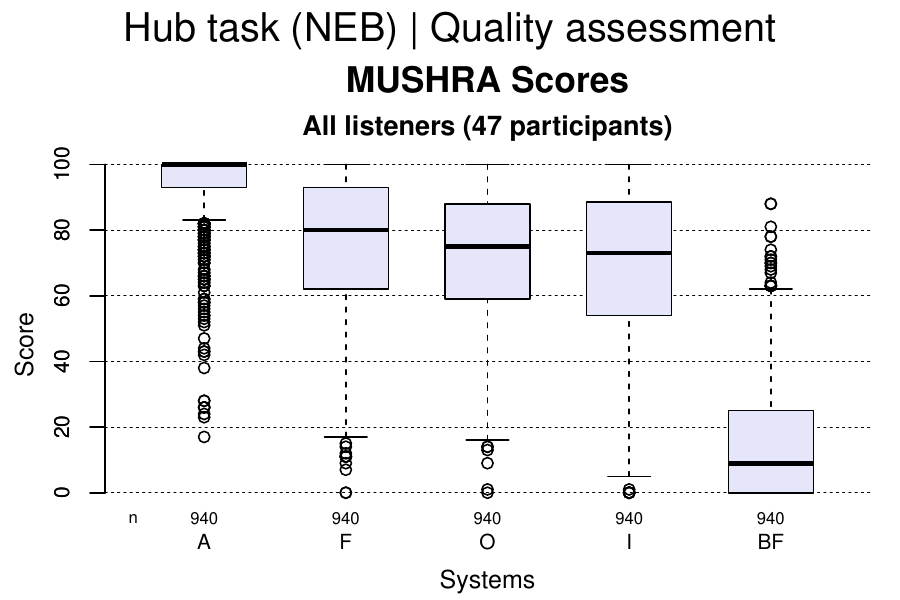}
  \caption{MUSHRA scores (all participants). Our System is O.}
  \label{fig:mushra_hub}
\end{figure}

The Table~\ref{tab:hub_eval} reports the initial evaluation scores of our system for the hub task, compared to the original recording. There were 361 to 228 validated participants for these initial tests which evaluated all 18 systems. The naturalness MOS of our system places it among the 3 best systems and close to natural recordings. In terms of speaker similarity, our system scored in the middle-range, on par with 10 other of the evaluated systems. However, the standard deviation of the similarity MOS is relatively high even for the natural recordings, which makes it difficult to interpret. Further assessment would be needed to understand if the difference is due to prosody (prominences, pauses) or speaker quality. The result of the intelligibility test shows that our system is robust to pronunciations errors and validates the unsupervised training approach of our non-attentive acoustic model. Finally, the results of the second evaluation round (MUSHRA test) are shown on Figure~\ref{fig:mushra_hub}. Three systems were evaluated (F, O, I) and our system ranked second, tied with the system I (no statistically significant differences, as shown in Blizzard detailed evaluations). 
\begin{table}[h!]
  \caption{Initial scores for our system compared to \textbf{(original)}.}
  \label{tab:hub_eval}
  \centering
  \begin{tabular}{lllll}
    \toprule
    Test (Hub task)    & Median & Mean & Sd & Metric    \\
    \midrule
    Naturalness   & 4 \textbf{(5)} & 4.2 \textbf{(4.4)} & 0.85 \textbf{(0.8)} & MOS   		\\
    Similarity    & 3 \textbf{(4)} & 2.9 \textbf{(3.4)} & 1.27 \textbf{(1.31)} & MOS          \\
    SUS Intelligibility & 0.0 & 0.11 & 0.2 & WER   \\
    \bottomrule
  \end{tabular}
\end{table}

\subsection{Spoke task}
\begin{figure}[h!]
  \centering
  \includegraphics[width=0.9\linewidth]{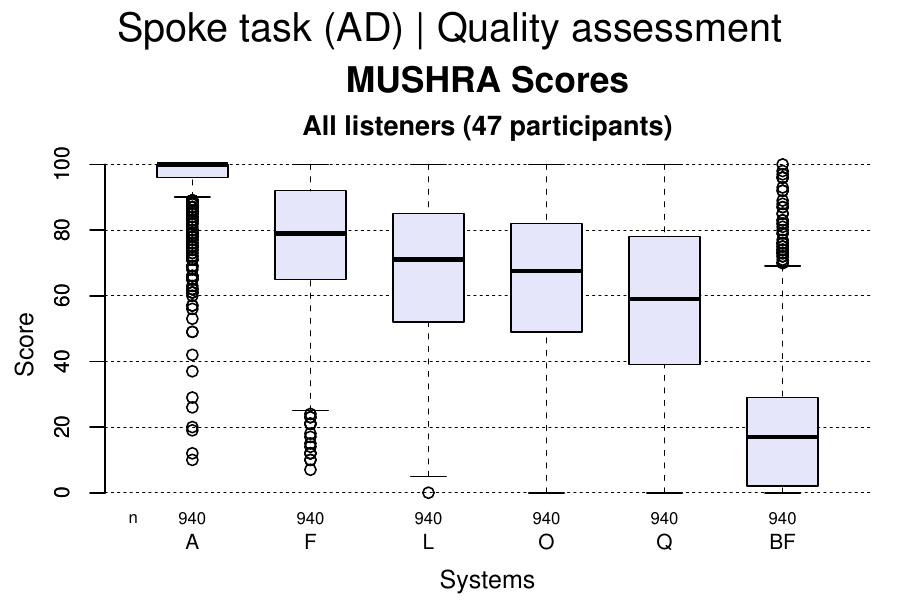}
  \caption{MUSHRA scores (all participants). Our System is O.}
  \label{fig:mushra_spoke}
\end{figure}

The Table~\ref{tab:spoke_eval} reports the initial evaluation scores of our system for the spoke task, compared to the original recording. There were 286 to 282 validated participants for these tests which evaluated all of the 14 systems for the spoke task. The naturalness MOS of our system places it among the 2 best systems, almost on par with natural recordings, although the "compressed scale" effect due low performing baseline should be accounted here. Like for the hub task, our system scored in the middle-range in terms of similarity scores, on par with 7 other systems but with a high variance (also noted for the natural speech). Our system was selected among the 4 bests models for the second evaluation test (MUSHRA) whose results are shown on Figure~\ref{fig:mushra_spoke}. Although our system appears third on the plot, there is no statistically significant difference with the second system L, as shown in the Blizzard detailed evaluations. Among the factors possibly limiting our system for the spoke task, it could be noted that there is a strong speaker imbalance in the dataset used to train the initial model (before fine-tuning). Also our speaker encoder was trained on 1400 speakers, which might not be enough to learn an exhaustive speaker embedding space.

\begin{table}[h!]
  \caption{Initial scores for our system compared to \textbf{(original)}.}
  \label{tab:spoke_eval}
  \centering
  \begin{tabular}{lllll}
    \toprule
    Test (Spoke task)   & Median & Mean & Sd & Metric    \\
    \midrule
    Naturalness   & 5 \textbf{(5)} & 4.4 \textbf{(4.5)} & 0.72 \textbf{(0.77)} & MOS   		\\
    Similarity    & 4 \textbf{(4)} & 3.4 \textbf{(4)} & 1.31 \textbf{(1.21)} & MOS          \\
    \bottomrule
  \end{tabular}
\end{table}

\section{Conclusions}
In this paper, we have presented our proposed two-stage TTS system for the Blizzard Challenge 2023. Our approach to achieve the best possible acoustic and prosodic naturalness relies on several key aspects. First, we choose an auto-regressive model for its high naturalness but use a duration predictor and gaussian upsampling at inference for better robustness. We devise a simple unsupervised training using gaussian attention. Second, we model the prosodic variability at both global and local level using a style token framework. Finally we use a BERT model to predict the global and local style embeddings, as well as pronunciation variants, conditioned on the speaker information. Our system was shortlisted for the second round of the Blizzard evaluation and performed second best in hub task and third best in spoke task. Further work would be to improve speaker adaptation and style transfer capabilities,
\bibliographystyle{IEEEtran}
\bibliography{mybib}

\end{document}